\documentclass[useAMS,usenatbib]{mn2e}

\usepackage[]{graphicx}

\newcommand{\x}{$\times$}
\newcommand{\ie}{{\emph i.e.~}}

\newcommand{\ptvp}{\emph{~ptv}\space}
\newcommand{\rms}{\emph{~rms}}

\newcommand{\mic}{\ensuremath{\mu}m}

\title[Radial thresholding to mitigate LGS aberrations]{Radial thresholding to mitigate Laser--Guide--Star aberrations on Centre--of--Gravity--based Shack--Hartmann wavefront sensors}

\author[Olivier Lardi\`ere, Rodolphe Conan, Colin Bradley, Kate Jackson and Peter Hampton]{Olivier Lardi\`ere\thanks{E-mail: lardiere@uvic.ca}, Rodolphe Conan, Colin Bradley, Kate Jackson and Peter Hampton\\
AO Laboratory, Mechanical Engineering Department, University of Victoria,\\ PO Box 3055 STN CSC, Victoria, BC, V8W 3P6, Canada}

\begin{document}

\date{Accepted ... . Received ... ; in original form ...}

\pagerange{\pageref{firstpage}--\pageref{lastpage}} \pubyear{2009}

\maketitle

\label{firstpage}

\begin{abstract}
Sodium Laser Guide Stars (LGSs) are elongated sources due to the thickness and the finite distance of the sodium layer. The fluctuations of the sodium layer altitude and atom density profile induce errors on centroid measurements of elongated spots, and generate spurious optical aberrations in closed--loop adaptive optics (AO) systems. According to an analytical model and experimental results obtained with the University of Victoria LGS bench demonstrator, one of the main origins of these aberrations, referred to as LGS aberrations, is not the Centre--of--Gravity (CoG) algorithm itself, but the thresholding applied on the pixels of the image prior to computing the spot centroids. A new thresholding method, termed ``radial thresholding'', is presented here, cancelling out most of the LGS aberrations without altering the centroid measurement accuracy.   
\end{abstract}

\begin{keywords} 
instrumentation: adaptive optics -- methods: analytical, laboratory.
\end{keywords}



\section{Introduction}
\label{sec:intro}

\begin{figure}
  \centering
  \includegraphics[width=\linewidth]{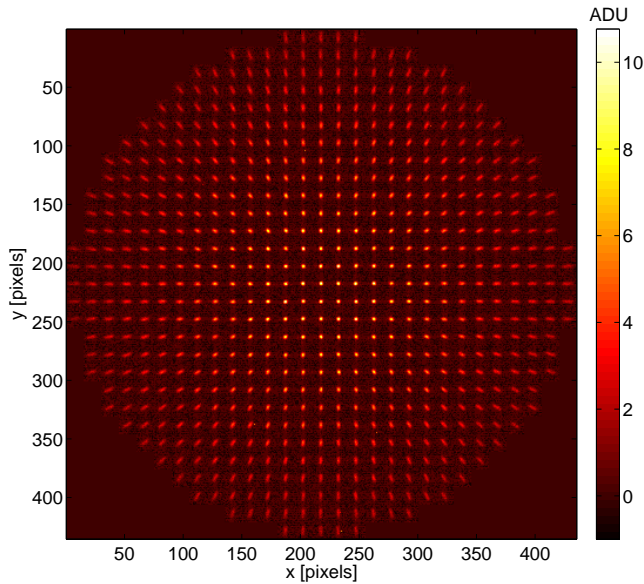}
  \caption{29\x29 elongated spot SH--WFS frame obtained with the UVic LGS--bench at SNR=18. The spot sampling is 2\x8~pixel on the edge pupil, corresponding to 1"\x4" on the sky, as expected for the TMT SH--WFS with a 1" seeing ($\sigma_{RON}$=0.43~ADU, no threshold).}  \label{fig:image}
\end{figure}

Sodium laser guide star (LGS) adaptive optics (AO) systems allow in theory a full sky coverage; however, there are several limitations. The artificial star is elongated due to the sodium layer thickness (about 10~km) and finite distance (90~km). Consequently, if the laser is launched from the secondary mirror holder, the spots of a Shack--Hartmann wavefront sensor (SH--WFS) are radially elongated from the centre to the edge of the pupil (Fig.~\ref{fig:image}). The spot elongation is proportional to the telescope diameter and can reach several arcseconds for extremely large telescopes (ELTs).

Moreover, the sodium layer is not static, but fluctuates with a time scale of about 1~minute or less~\citep{davis}. Fluctuations of the sodium layer altitude and atom density profile induce errors on centroid measurements of elongated spots and generate spurious aberrations on the wavefront in closed--loop AO systems. These aberrations, referred to as LGS aberrations, can reach several hundred nanometres peak--to--valley (\emph{ptv}) for ELTs with the classical centre--of--gravity (CoG) centroiding algorithm~\citep{clare, lardiere}. Some authors proposed new sophisticated centroiding algorithms to mitigate LGS aberrations, such as the Matched Filtering~\citep{gilles, conan} or the correlation~\citep{pyoneer, thomas, thomas08}. 

However, thanks to the LGS--bench demonstrator built at the University of Victoria (UVic) for ELT projects~\citep{lardiere, conan}, we found out that one of the main sources of the LGS aberrations was not the CoG algorithm itself, but simply the threshold applied on the pixels of the SH--WFS images before the centroid computation. 

Section~\ref{sec:origins} reviews the known possible origins of LGS aberrations. A model of the aberrations generated by thresholding is presented in Sec.~\ref{sec:model}, as well as a simple new thresholding method, termed ``radial thresholding'', which mitigates most of LGS aberrations. The experimental results obtained with the UVic bench with the radial thresholding show that the CoG algorithm is still well suited for LGS wavefront sensing on ELTs (Sec.~\ref{sec:results}).

\section{Origins of LGS aberrations}
\label{sec:origins} 

If the LGS spots are radially elongated from the pupil centre, as shown on Fig.~\ref{fig:image}, the sodium layer fluctuations induce centro--symmetric aberrations, such as focus ($Z_4$) and spherical aberrations ($Z_{11}$, $Z_{22}$, etc.), and also square symmetric aberrations, such as tetrafoils ($Z_{14}$, $Z_{26}$, etc.). 

The focus is due to a variation of the sodium layer altitude. This error is not an artefact and must be compensated by refocussing the LGS on the WFS with zoom optics and by updating the offsets of the LGS WFS with a natural--guide--star (NGS) focus sensor~\citep{herriot}. Non--common path errors of the LGS optical train, including the zoom optics, can vary with the sodium layer distance, \ie the zenithal angle, and induce variable aberrations on the science path too. We assume that these systematic aberrations can be calibrated and virtually negated.

Consequently, aberrations beyond focus are mainly artefacts of the wavefront sensing. According to a model from \citet{clare} and to the first experimental results obtained with the UVic LGS bench~\citep{lardiere}, the spherical aberrations arise due to a truncation of asymmetric LGS spots by a circular field-–stop, while square symmetric aberrations are likely due to:
\begin{itemize}
\item a spot truncation by a square field–-stop or by pixel boundaries, horizontally and vertically elongated spots being more truncated than diagonally elongated spots,
\item a spot overlap for square--grid lenslet arrays, horizontally and vertically elongated spots being more prone to overlapping,
\item quad–-cell or sampling effects on centroid measurements.
\end{itemize}  

Both kinds of LGS aberrations have been reproduced and characterized in laboratory on the UVic bench with a time--series of 88 real sodium profiles (Fig.~\ref{fig:Naseq}). Beyond the focus, the most significative LGS aberrations detected are the spherical aberration $Z_{11}$ up to 100nm\ptvp (30nm\rms), and the tetrafoil $Z_{14}$ with 40nm\ptvp (10nm\rms). Moreover, a correlation between the spherical aberration and the profile asymmetry was empirically established~\citep{lardiere}. Square--symmetric aberrations, such as $Z_{14}$, should be mitigated by using a polar--coordinate CCD array \citep{beletic, thomas08}.  

However, we discovered later that $Z_{11}$ mode disappears if no threshold was applied on the pixels of the LGS WFS images before the computation of centroids. The thresholding discards the two extremities of each elongated spot, and consequently truncates radially each spot, as an optical circular field--stop would do. The spot truncation caused by the field--stop is negligible compared to the truncation induced by the pixel thresholding if the field of view (FOV) of the outermost lenslets is wide enough to make an image of a 20~km--thick sodium profile.

With such a large FOV and a polar--coordinate CCD array, as expected for the Thirty--Meter--Telescope (TMT) LGS--AO facility \citep{tmt}, the thresholding is likely the main source of the LGS aberrations and deserves a specific study.  

\section{Modelling the aberrations induced by thresholding}
\label{sec:model}
Basically, a thresholding must be applied on the image pixels prior to computing the spot centroids in order to minimize the contribution of the detector read--out noise $\sigma_{RON}$, or of the sky background. The thresholding is generally uniform over the pupil and is implemented as follow:
\begin{equation}
I_t(x,y) = \left\{ \begin{array}{rcl}
I(x,y)-\mbox{\emph{Thres}} & \mbox{if} & I(x,y)\geq \mbox{\emph{Thres}} \\
0 & \mbox{if} & I(x,y)<\mbox{\emph{Thres}}
\end{array}\right. ,
\label{equ:It}
\end{equation}
\noindent with $I$ and $I_t$ the raw and the thresholded images respectively. \emph{Thres} is the intensity level of the threshold expressed in detector counts, \ie in Analog--to--Digital Units (ADU). Generally, the threshold is defined from the readout noise (at $3\,\sigma_{RON}$ for instance). The remainder of this section demonstrates how a uniform thresholding can induce spherical aberrations when the spot profile is asymmetric.

\subsection{Geometrical model of the spot elongation}

\begin{figure}
  \centering
  \includegraphics[width=0.85\linewidth]{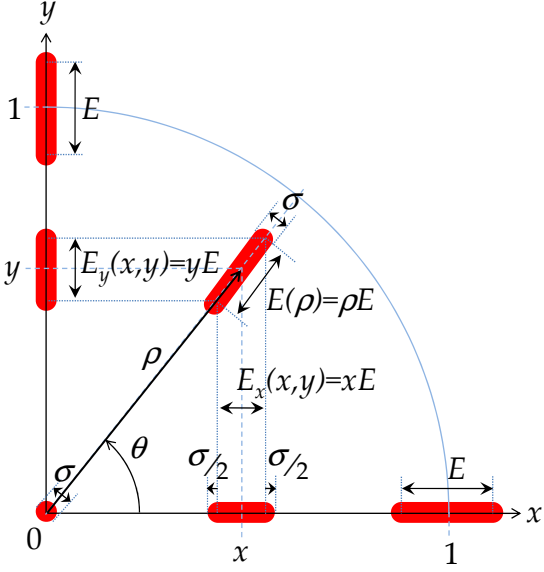}
  \caption{Geometrical model of the spot elongation if the laser is launched from the centre of the pupil.}
  \label{fig:geo}
\end{figure}

If the sodium laser is launched on the system optical axis, the spots of a SH--WFS are radially elongated from the pupil centre, as shown on Fig.~\ref{fig:geo}. Let $\rho$ be the distance of a lenslet to the pupil centre, normalized to $1$ on the pupil edge. From Eq.~1 of \citet{lardiere}, the geometrical radial size, on the detector plane, of the corresponding spot is proportional to $\rho$, such as $E(\rho)\approx E\,\rho$, with
\begin{equation}
E=\frac{f\,R\,\sigma_{Na}}{h_0^2}\,cos(z)\,.
\end{equation}

In this equation, $h_0$ and $\sigma_{Na}$ are the mean altitude and the thickness of the sodium layer respectively, $z$ is the zenithal angle, $R$ is the radius of the telescope pupil, and $f$ is the resultant focal length of the whole optical system. $E$ is the geometrical radial extent, on the detector plane, of a spot located on the edge of the pupil (\ie $\rho=1$). $E(\rho)$ is simply termed spot elongation hereafter, and can be expressed in $\mu$m or in pixels.

If $x$ and $y$ are the Cartesian coordinates of a spot in the detector plane, then the geometrical sizes of that spot, projected on the $x$ and $y$ axis are respectively defined as follow:

\begin{equation}
\left\{
\begin{array}{l}
E_x(x,y) = x\,E\\
E_y(x,y) = y\,E\\
\end{array}
\right.
\label{equ:Exy}
\end{equation} 

Let $I_o$ and $\sigma$ be the maximum intensity and the size of the central spot (\ie $\rho=0$) respectively. The extent $\sigma$, expressed in the same unit than $E$, is defined by the blurring due to the lenslet diffraction lobe, or to the seeing. Considering this blurring effect, the actual length and width of a spot located at the radius $\rho$ are respectively $E(\rho)+\sigma$ and $\sigma$. Let $I_{max}(\rho)$ be the maximum intensity of that spot. As the flux is conserved for all spots of the pupil, we have the following equation:
\begin{equation}
\sigma^2I_o\approx \sigma\,(E(\rho)+\sigma)\,I_{max}(\rho)\,.
\end{equation}
Hence,
\begin{equation}
I_{max}(\rho)\approx\frac{I_o}{1+\frac{E}{\sigma}\rho}\,.
\label{equ:Imax}
\end{equation}

\subsection{Centroid errors versus threshold}

Let $P_0(r)$ be the projection of the sodium layer vertical profile on the SH--WFS detector plane, for a lenslet located on the edge of the pupil (\ie $\rho=1$). This projection is not linear and takes in account the perspective effect which stretches the lower part of the sodium layer and compress the upper part. Thus, the one--dimensional function $P_0(r)$ is the radial intensity profile of the geometrical image of the most elongated spot. The variable $r$ is the radius position defined inside the sub--image plane of the considered spot only. If a threshold of value $Thres$ is applied on the spot image, the thresholded profile, obtained accordingly to Eq.~\ref{equ:It}, is noted $P_t(r)$, with $t=Thres/I_{max}(\rho)$ the threshold value normalized to the local spot maximum intensity. For any spot profile and any threshold $t\in[0\,;1]$, we can define the dimension--less function $g(t)$ such as
\begin{equation}
g(t)=\frac{1}{E}\left( \frac{\int r\,P_t(r)\,dr}{\int P_t(r)\,dr} - \frac{\int r\,P_0(r)\,dr}{\int P_0(r)\,dr}\right)\,.
\label{equ:gt}
\end{equation}

\begin{figure}
  \centering
  \includegraphics[width=\linewidth]{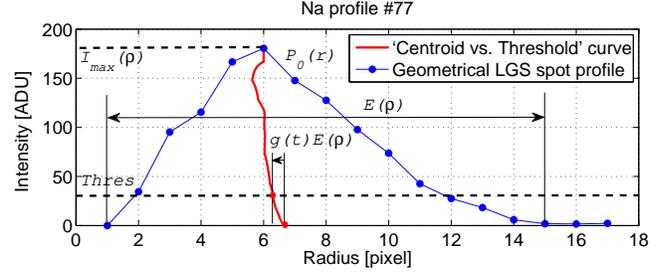}
  \caption{Radial profile of the theoretical geometrical image of the most elongated spot for a sodium profile featuring a strong asymmetry (profile \#77 of the sequence displayed in Fig.~\ref{fig:Naseq}). This one--dimensional curve is the projection of the vertical sodium profile on the SH--WFS detector plane, sampled on a finite number of pixels, with no diffraction, no atmospheric turbulence, no photon noise and no read--out noise.}
  \label{fig:gt}
\end{figure}

According to our previous assumption, this function provides, for any spot of the pupil, the radial centroid error induced by the threshold $t$, expressed as a fraction of the elongation of the considered spot. In absolute units (pixel or $\mu$m), the radial centroid error scales as the spot elongation, and is $g(t)\,E(\rho)$. We note that $g(t)=0$ for symmetric profiles. Figure~\ref{fig:gt} plots a geometrical spot profile $P_0(r)$ with its function $g(t)$. This profile is among the most asymmetric sodium profiles found in the 88--profile sequence of Fig.~\ref{fig:Naseq}. The projection of the radial error $g(t)\,E(\rho)$ on the $x$ and $y$--axis, provides $\delta_x(x,y)$ and $\delta_y(x,y)$, the errors made on the wavefront slopes in $x$ and $y$--directions respectively for a spot located in coordinates ($x$,$y$):

\begin{equation}
\left\{
\begin{array}{l}
\delta_x(x,y) = E_x(x,y)\,g(t)=x\,E\,g(t)\\[1ex]
\delta_y(x,y) = E_y(x,y)\,g(t)=y\,E\,g(t)\\
\end{array}
\right.
\label{equ:deltaxy}
\end{equation}  

Hence, the wavefront error $Z(x,y)$ caused by thresholding is deduced from the slopes, such as, 
\begin{equation}
\left\{
\begin{array}{l}
\frac{\partial Z}{\partial x}= x\,E\, g(t)\\[1ex]
\frac{\partial Z}{\partial y}= y\,E\, g(t)\\
\end{array}
\right.
\label{equ:xy}
\end{equation}  

Using a polar coordinate system ($\rho$, $\theta$), such as $x=\rho \cos\theta$ and $y=\rho \sin\theta$, we can use the following conversion formulae,
    
\begin{equation}
\left\{
\begin{array}{l}
\rho\frac{\partial Z}{\partial \rho}=x\frac{\partial Z}{\partial x} + y\frac{\partial Z}{\partial y}\\[1ex]
\frac{\partial Z}{\partial \theta}=-y\frac{\partial Z}{\partial x} + x\frac{\partial Z}{\partial y}\\
\end{array}
\right.
\label{equ:xyrt}
\end{equation} 

\noindent to obtain,
\begin{equation}
\left\{
\begin{array}{l}
\frac{\partial Z}{\partial \rho}= \rho \,E\,g(t)\,,\\[1ex]
\frac{\partial Z}{\partial \theta}=0\,.\\
\end{array}
\right.
\label{equ:rt}
\end{equation}

Thus, the wavefront distortion generated by thresholding is always a surface of revolution whatever $g(t)$, \ie whatever the sodium profile. In terms of Zernike polynomials~\citep{noll}, this result proves that the thresholding can only induce a combination of focus ($Z_4$) and spherical aberrations ($Z_{11}$, $Z_{22}$, $Z_{37}$, etc.).  

\subsection{Wavefront aberrations induced by uniform thresholding}

If the threshold value $Thres$ is constant for the whole pupil, then $t=\mbox{\emph{Thres}}/I_{max}(\rho)$ varies with $\rho$. From Eqs.~\ref{equ:Imax} and~\ref{equ:rt}, we obtain in that case:
\begin{equation}
\frac{\partial Z}{\partial \rho}\approx \rho\,E\,g\left(\frac{\mbox{\emph{Thres}}}{I_o}\left(1+\frac{E}{\sigma}\rho\right)\right)\,.
\label{equ:dZ}
\end{equation}  

Moreover, we assume that $g(t)$ can be expressed, otherwise approximated, as a polynomial of degree $n$ such as:
\begin{equation}
g(t)\approx\sum_{i=1}^n g_i\,t^i
\end{equation}
 
\noindent with the sum starting at $i=1$ because $g(0)=0$ by definition~(Eq.~\ref{equ:gt}). Using the binomial theorem, we can find, after integration of Eq.~\ref{equ:dZ}:


\begin{eqnarray}
Z(\rho)\approx \hspace{.74\linewidth} \nonumber \\ 
E\,\sum_{i=1}^n g_i\,\left(\frac{\mbox{\emph{Thres}}}{I_o}\right)^i\,
\sum_{j=0}^i\frac{1}{j+2}\frac{i!}{j!(i-j)!}\left(\frac{E}{\sigma}\right)^j\rho^{j+2}\,. 
\label{equ:Z}
\end{eqnarray}

We can check that $Z(\rho)=0$ if the spots are not elongated ($E=0$), or if the threshold is null, or again if the spot profile is symmetric ($g_i=0\,,\forall i$). If $n$ is the order of the polynomial $g(t)$, the radial order of the wavefront aberration is $n+2$. A uniform threshold applied on a triangular or trapezoidal (\ie $n=1$) asymmetric spot profile induces already aberrations of the third radial order:
\begin{equation}
Z(\rho)\approx \frac{E}{2}\,g_1\,\frac{\mbox{\emph{Thres}}}{I_o}\,\left(\rho^2+\frac{2}{3}\frac{E}{\sigma}\rho^3\right)\,.
\label{equ:Ztri}
\end{equation}

The same threshold applied on a more complex asymmetric sodium profile will induce higher order aberrations on the wavefront. The projection of $\rho^2$ on the Zernike polynomials base is a pure focus:
\begin{equation}
\rho^2=\frac{1}{2\sqrt{3}}\,Z_4\,,
\label{equ:rho2}
\end{equation}
\noindent while $\rho^3$ is a combination of focus and all spherical aberrations ($Z_{11}$, $Z_{22}$, $Z_{37}$, etc.):  
 \begin{equation}
\rho^3\approx 0.3\,\left(Z_4+0.14\,Z_{11}-0.015\,Z_{22}+0.003\,Z_{37}+...\right)
\label{equ:rho3}
\end{equation}

The focus term is corrected by the NGS focus sensor already required to offset the altitude variation of the sodium layer. Beyond the focus, the main aberration induced by a uniform thresholding is $Z_{11}$ mode, higher order spherical aberrations being negligible. The $Z_{11}$ error is fluctuating with the time, due to the variations of the sodium profile asymmetry, and is an issue for LGS AO systems.
    
\subsection{Wavefront aberrations induced by radial thresholding}

If the threshold value is now, not uniform over the whole pupil, but defined for each sub--image proportionally to the maximum intensity of the local spot, then the normalized threshold $t=\mbox{\emph{Thres}}/I_{max}(\rho)$ becomes constant, as well as $g(t)$. From Eq.~\ref{equ:rt}, we deduce that
\begin{equation}
Z(\rho)=\frac{E}{2}\,g(t)\,\rho^2\,.
\label{equ:Zrho2}
\end{equation}
  
Consequently, such a thresholding method induces only a focus error and no spherical aberrations at all, whatever the sodium profile structure. As the threshold value scales as the inverse of the radius, this thresholding is refereed to as ``radial thresholding''. 

\section{Laboratory results of the radial thresholding}
\label{sec:results}

\begin{figure}
  \centering
  \includegraphics[width=\linewidth]{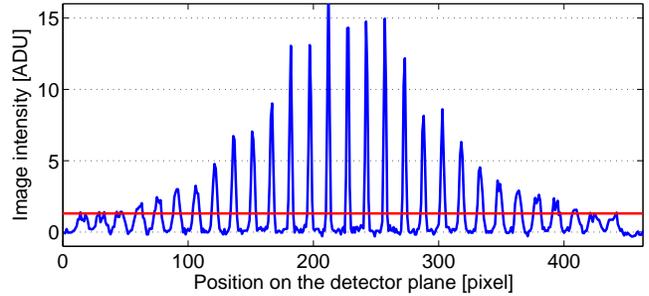}
  \caption{Horizontal radial cut of the 29\x29 elongated spot SH--WFS frame of Fig.~\ref{fig:image} obtained with the UVic LGS--bench at SNR=18. The red solid line illustrates a uniform threshold set at 3 times the read--out noise ($\sigma_{RON}$=0.43~ADU on the bench). The radial cut is a binning of the 3 central pixel lines of the frame. According to the spot elongation, the theoretical intensity ratio between the central spots and the edge spots is 4. The actual ratio is around 7 here due to the gaussian profile of the collimated beam used on the bench, and also to the field curvature of the imaging lens. These effects exaggerate slightly the wavefront errors induced by thresholding.}
  \label{fig:hcut}
\end{figure}

\begin{figure}
  \centering
  \includegraphics[width=\linewidth]{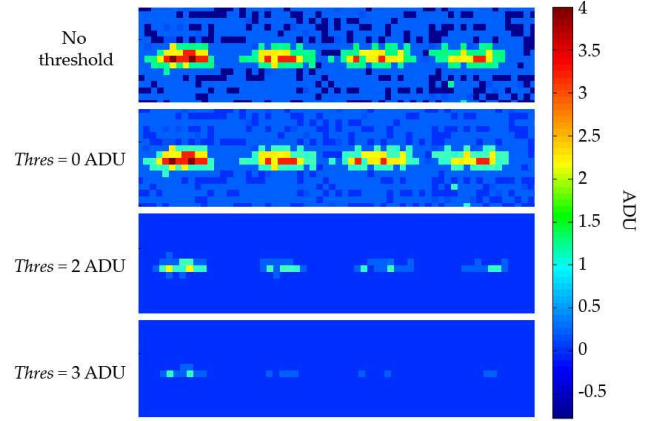}
  \caption{Truncation effect induced by thresholding on the 4 outer spots at SNR=18, for 3 different threshold levels. The 0--ADU level is the detector offset. Pixel values might be negative due to read--out noise.}
  \label{fig:zoom}
\end{figure}

\subsection{Experimental setup}
The radial thresholding is easy to implement on real SH--WFSs, and has been validated in laboratory with the UVic LGS demonstrator bench and compared to the results obtained without threshold or with an uniform threshold. Two signal--to--noise ratios (SNR) have been considered, 140 and 18, the latter being expected for real LGSs. At SNR=140, the read--out noise is negligible, and thresholding could be not necessary. However, this favourable case is useful to properly distinguish the thresholding effect from the noise, and to validate our model. 

The readout noise of the SH--WFS camera is 0.43~ADU\rms. As the offset frame of the detector is subtracted to the raw images, some pixels may have negative values due to the read--out noise. Consequently, a 0--ADU threshold discards the negative values, and does not mean that no threshold is applied on the image. We will have to distinguish both cases in our following experimental study at low SNR.

The FOV of the circular field--stop of the bench corresponds to a sodium layer thickness of 20~km for the most elongated spot. As the mean thickness of the sodium layer is about 10~km, the spot truncation effect due to the field--stop should be negligible compared to the spot truncation caused by the thresholding.

Figure~\ref{fig:image} displays a 29\x29 radially elongated spot image obtained with the UVic bench for SNR=18. On this image, the maximum spot elongation $E$ is 8 pixels, and the spot width $\sigma$ is 2 pixels. Figure~\ref{fig:hcut} plots an horizontal cut of this image ($y=0$). A uniform threshold at 1.3~ADU, \ie $3\,\sigma_{RON}$ for the bench, is shown as an example. This graph already points up that a uniform thresholding is not suited for LGS spots, since edge spots are fainter than the central spots. At low SNR, a uniform threshold might even reach the maximum intensity of some spots located on the edge. Figure~\ref{fig:zoom} illustrates this issue for SNR=18. 

\begin{figure}
  \centering
  \includegraphics[width=\linewidth]{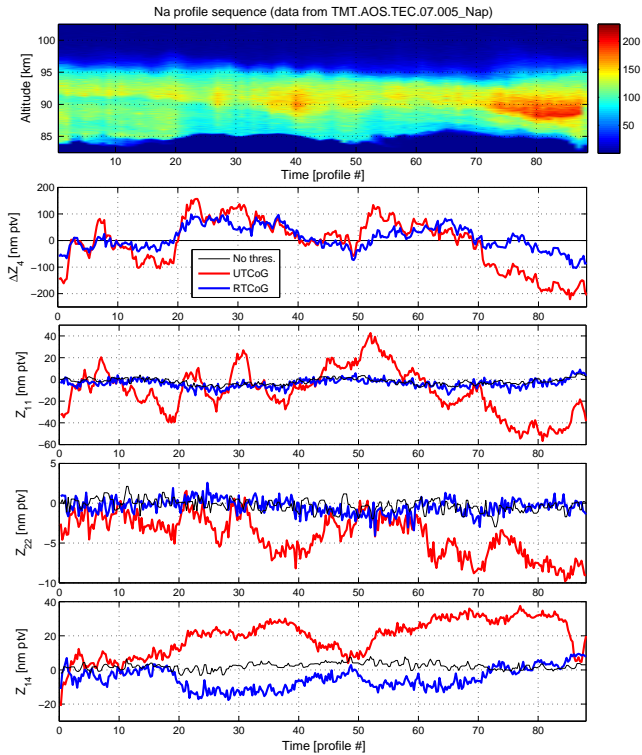}
  \caption{Sodium profile sequence used on the UVic LGS--bench (real data from Purple Crow LIDAR, Univ. of Western Ontario, altitude\x time resolution is 25~m\x70~s) and LGS aberrations measured by a centre--of--gravity--based SH--WFS in open--loop with no turbulence, when no threshold, or a uniform threshold (UTCoG), or again a radial threshold (RTCoG) is applied on the pixels of LGS spot images. The focus ($Z_4$) component due to an altitude change of the sodium layer is removed here to highlight the component due to the thresholding (SNR=140, UT=13~ADU, RT=4\% of the local spot maximum, $\sigma_{RON}$=0.43~ADU).}  \label{fig:Naseq}
\end{figure}

\begin{figure}
  \centering
  \includegraphics[width=\linewidth,clip]{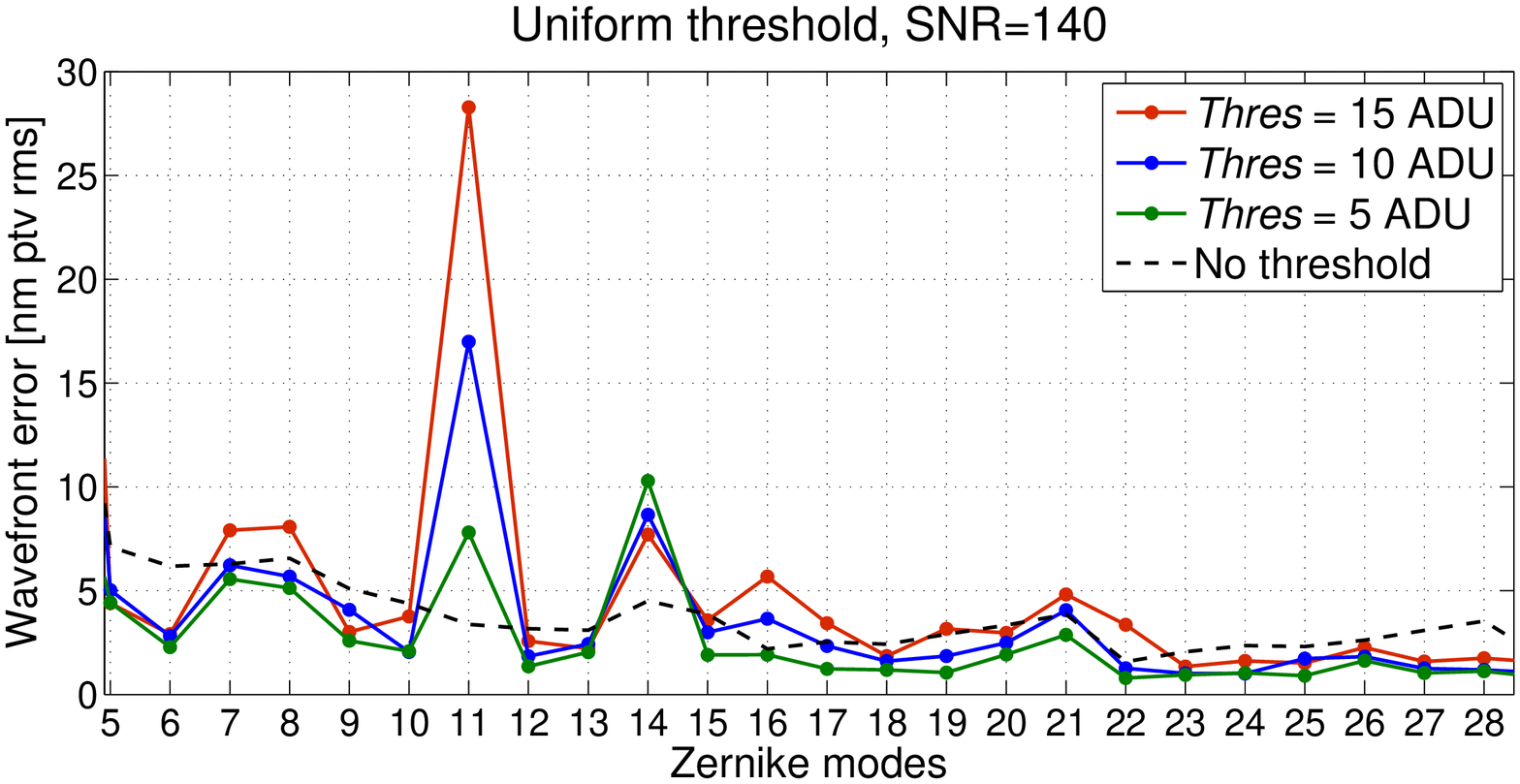}
\begin{center} (a) \end{center}
\vspace{4mm}
\includegraphics[width=\linewidth,clip]{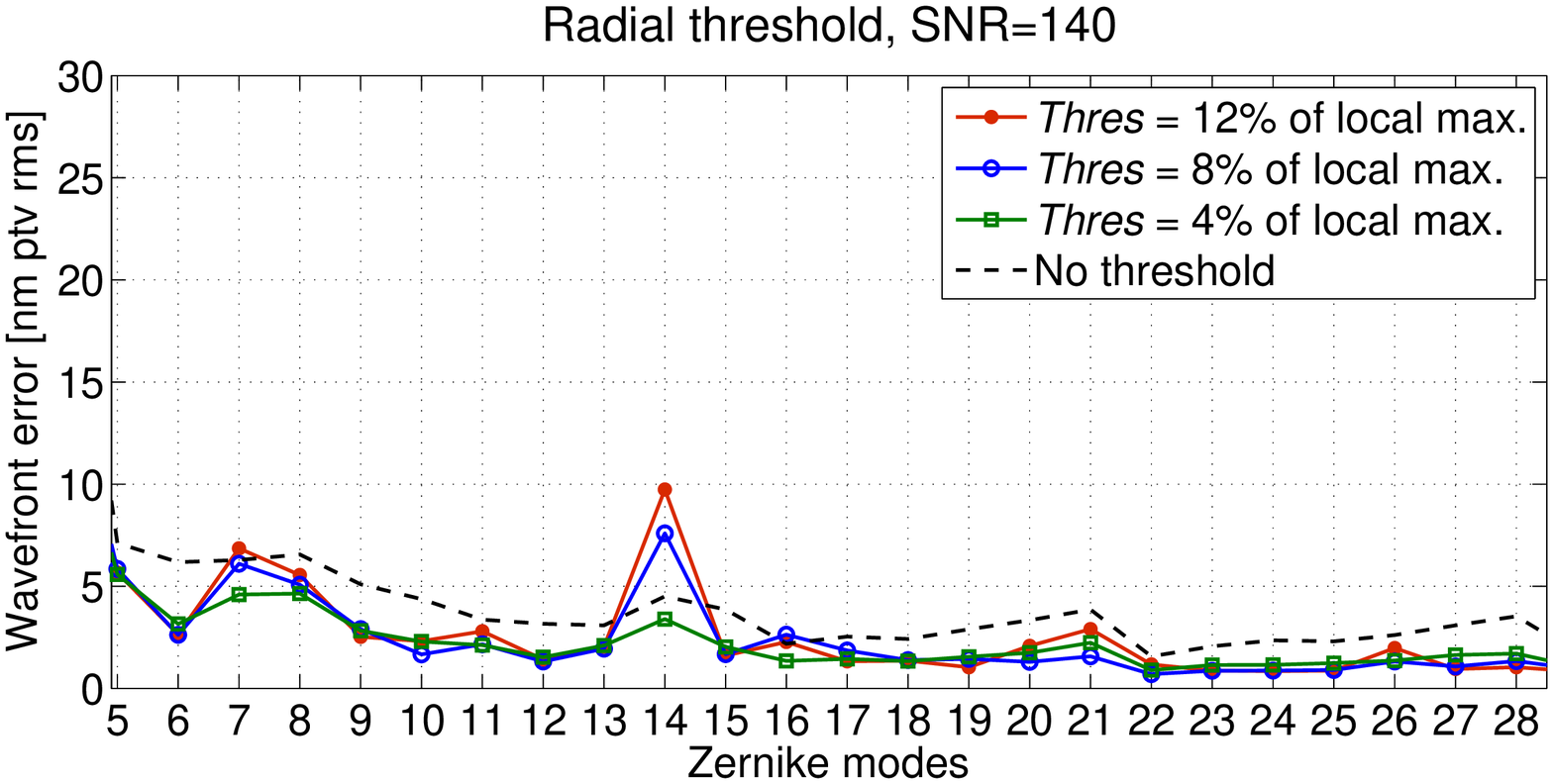}
\begin{center} (b) \end{center}
  \caption{LGS--induced aberrations measured on the UVic LGS--bench at SNR=140 when a uniform threshold (a) or a radial threshold (b) is applied on LGS spot images ($\sigma_{RON}$=0.43~ADU).}
  \label{fig:SNR140}
\end{figure}

\begin{figure}
  \centering
  \includegraphics[width=\linewidth,clip]{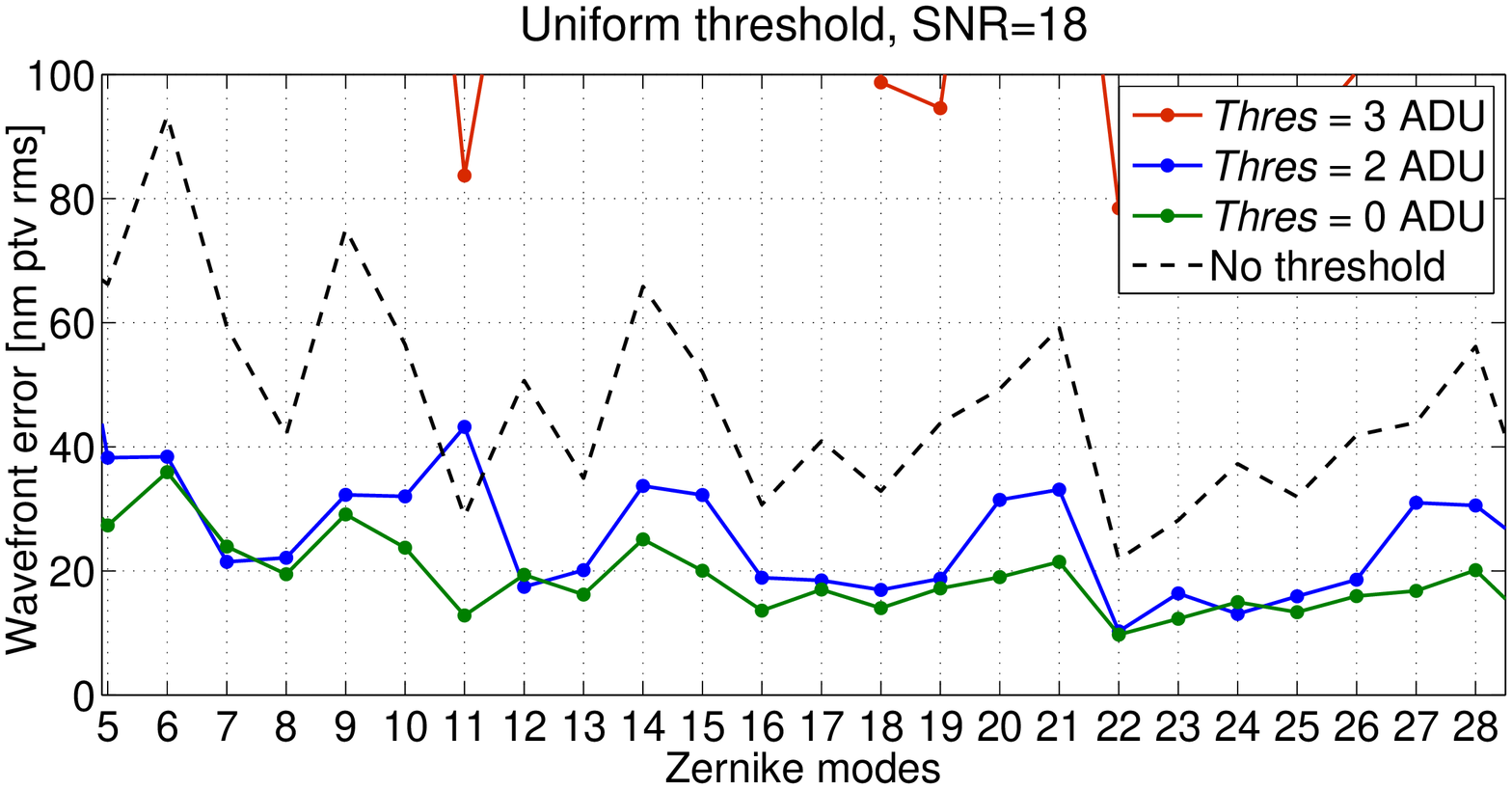}
\begin{center} (a) \end{center}
\vspace{4mm}
  \includegraphics[width=\linewidth,clip]{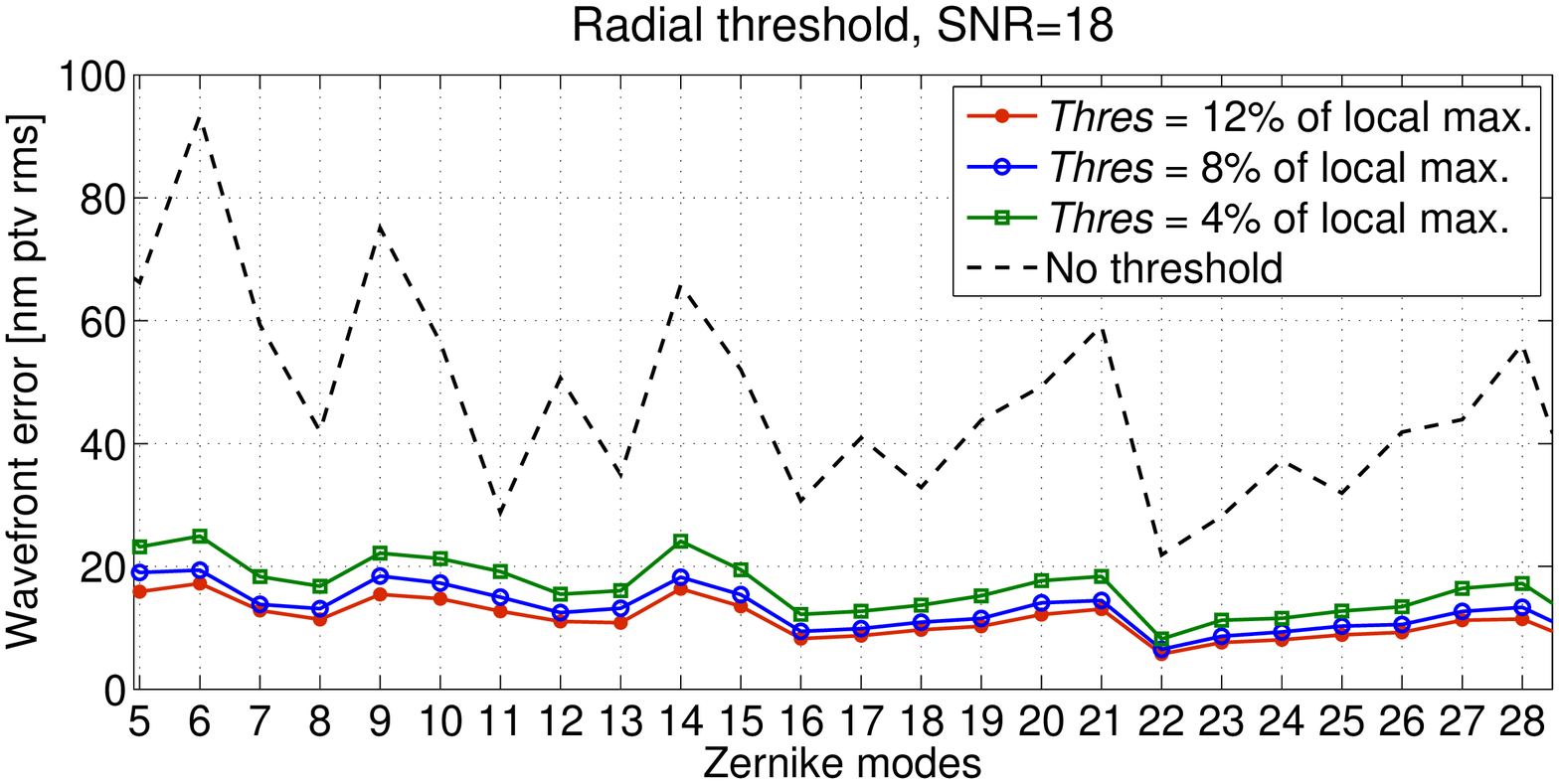}
\begin{center} (b) \end{center}
  \caption{LGS--induced aberrations measured on the UVic LGS--bench at SNR=18 when a uniform threshold (a) or a radial threshold (b) is applied prior to computing centroids ($\sigma_{RON}$=0.43~ADU).}
  \label{fig:SNR18}
\end{figure}

\subsection{Performance at high flux} 

Figure~\ref{fig:Naseq} plots the LGS aberrations for $Z_4$, $Z_{11}$, $Z_{22}$ and $Z_{14}$ modes versus the time, measured during the whole sodium profile sequence, with no threshold, with a 13--ADU uniform threshold and a radial threshold (4\% of local maximum) at SNR=140, in open--loop without turbulence. Such a high uniform threshold, not suited for real AO systems, is chosen here to exaggerate the thresholding effects on the wavefront and validate the model. Figure~\ref{fig:SNR140} plots the corresponding temporal RMS fluctuation of all Zernike modes beyond focus up to $Z_{28}$, for SNR=140 and for different threshold values (both uniform and radial).

As the main focus error source is the altitude variation of the sodium layer, the focus obtained without thresholding has been subtracted from those obtained with thresholding. This subtraction highlights the extra focus error induced by thresholding. This error reaches $350$nm\ptvp with the uniform threshold, and 200nm\ptvp with the radial threshold. Moreover, a correlation is visible between the focus and $Z_{11}$ modes induced by the uniform thresholding. This correlation confirms that $Z_4$ and $Z_{11}$ modes arise together. This result is in good agreement with the model (Eqs.~\ref{equ:Z} to~\ref{equ:rho3}). Although the ratio $Z_4/Z_{11}$ depends on the coefficients $g_i$, \ie the sodium profile structure and asymmetry, we can attempt to quantify it for a linear fit of the functions $g(t)$, \ie for a sequence of trapezoidal or triangular sodium profiles. In this simple case, the ratio $Z_4/Z_{11}$ expected by the model is about $7.3$ with $E/\sigma=4$ (Eqs.~\ref{equ:Ztri} to~\ref{equ:rho3}), while the observed ratio is about $4.9$ for the\ptvp amplitudes (Fig.~\ref{fig:Naseq}). The model and the experiments provide the same order of magnitude for the ratio $Z_4/Z_{11}$. The difference is likely due to the assumption made about the profile shape. A higher order fit of the functions $g(t)$ would be required for each profile, as well as further measurements using a symmetric sodium profile as a reference.  

Figure~\ref{fig:Naseq} shows clearly that the radial thresholding mitigates the spherical aberrations ($Z_{11}$ and $Z_{22}$) induced by the sodium layer variations, as if no thresholding was applied. The variations for the $Z_{11}$ mode reaches 90nm\ptvp with uniform thresholding, against 10nm\ptvp with radial thresholding or without thresholding at all. The remaining error is likely the component induced by the noise and by the circular field--stop. Moreover, Fig.~\ref{fig:SNR140}a shows that the RMS fluctuation of $Z_{11}$ mode scales as the level of the uniform threshold, which is in agreement with the model too (Eq.~\ref{equ:Z}).

As expected, the radial thresholding does not mitigate the square symmetric LGS aberration $Z_{14}$. However, the $Z_{14}$ error is modified by the thresholding, and is almost null without thresholding (Fig.~\ref{fig:Naseq} and ~\ref{fig:SNR140}a). Hence, the thresholding would likely induce other effects not considered in our model. The pixel boundaries or the spot overlap may have more or less impact depending on the threshold value. Assuming that polar coordinate CCD arrays should mitigate square symmetric modes, we will not investigate with more details the other possible sources of this mode in this paper. 
 
At SNR=140, the best performances are obtained with a radial threshold set at 4\% of the local spot maximum. The RMS wavefront errors averaged over the pupil for all modes beyond the focus are summarized in Table~\ref{tab}.

\begin{table}
\caption{RMS wavefront errors (WFE), in nm\rms, and achievable Strehl ratios (S), in \% for $\lambda=1\mic$, obtained at SNR=140 and 18, during the sodium profile sequence with the center--of--gravity algorithm without pixel thresholding, with uniform thresholding, and with radial thresholding. The threshold levels are optimized for SNR=18. The uniform threshold values are 13 and 0~ADU for SNR=140 and 18 respectively, while the radial threshold values are 4\% and 12\% for SNR=140 and 18 respectively. All Zernike modes beyond the focus (from $Z_5$ to $Z_{28}$) are added up and spatially averaged over the pupil. The bench accuracy is the ultimate centroiding performance achieved by the bench on non--elongated spots. Readout noise is 0.43~ADU\rms. }
\vspace{1mm}
\centering
\begin{tabular*}{\linewidth}{@{\extracolsep{\fill}}l| c c|c c}
\hline
&\multicolumn{2}{c|}{\emph{SNR=140}}&\multicolumn{2}{c}{\emph{SNR=18}}\\
&\emph{WFE}&$S$&\emph{WFE}&$S$\\ \hline
No threshold&4.34&99.9&60.87&86.4\\
Uniform threshold&10.65&99.5&31.91&96.1\\
Radial threshold&4.27&99.9&18.70&98.6\\
\hline 
Bench accuracy&2.73&99.9&17.84&98.7\\[0.5ex]
\hline   
\end{tabular*}
\label{tab}     
\end{table}

\subsection{Performance at low flux} 

Figure~\ref{fig:SNR18} plots the temporal RMS fluctuation of Zernike modes from $Z_5$ to $Z_{28}$, for SNR=18 and for different threshold values, both uniform and radial.

At low SNR, edge spots are very faint, a few ADU. Consequently a uniform threshold, even low, can easily exceed the maximum of some spots, and induce huge errors. The errors induced by a 3--ADU threshold exceed 100~nm, the upper limits of the graph (Fig.~\ref{fig:SNR18}a). Hence, a small change of the threshold level or of the SNR, which is fluctuating with LGS, can decrease significantly the performances. A 0--ADU uniform threshold appears to be more robust in that case, but it is not optimal in term of noise.

The performance obtained with the radial threshold is more robust at SNR=18, and does not depend much on the threshold value (Fig.~\ref{fig:SNR18}b). This is certainly due to the ability of the radial thresholding to track the temporal  variations of the spot intensity: the radial thresholding is also a dynamic thresholding. The wavefront errors measured with a radial threshold are lower than those measured with a uniform threshold in the best case (Tab.~\ref{tab}). At SNR=18, the optimal value for the radial threshold is 12\% of the local maximum.

Lastly, Tab.~\ref{tab} shows that the performance achieved with the radial thresholding on elongated spots is very close to the ultimate performance achievable on the bench with a non--elongated source. This result basically means that the radial thresholding cancels out most of the errors induced by the LGS spot elongation, without altering the centroid accuracy in presence of noise.


\section{Discussions}
\subsection{Weighted Centre--of--Gravity algorithm}
Considering the issues caused by the thresholding on LGS spots, we could suggest to use the Weighted Centre--of--Gravity algorithm, which does not require a prior thresholding, instead of the conventional centre--of--gravity algorithm. Basically, this algorithm consists in weighting the pixels of the spot image by a reference image of the spot~\citep{nicolle}. If the spot image matches perfectly the reference image, this technique is equivalent to compute the center--of--gravity of the square of the image intensity. However, the centre of gravity of the square of an elongated spot image differs from the actual center of gravity of that spot, if the spot profile is asymmetric. Consequently, the Weighted--Center--of--Gravity algorithm cannot be used on LGS AO systems as it is.
  
\subsection{Case study of non--radial spot elongation}

\begin{figure}
  \centering
  \includegraphics[width=3.25in, clip]{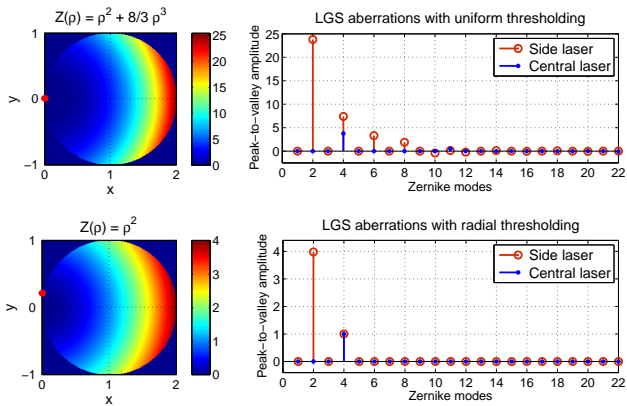}
  \caption{Typical LGS aberrations arising when the laser is launched from the side of the pupil (x=0 and y=0) with a uniform thresholding (top) and for a radial thresholding (bottom). These aberrations are compared, in term of Zernike modes, to those arising when the laser is centrally projected. Aberration amplitudes are peak--to--valley in an arbitrary unit.}
  \label{fig:offaxis}
\end{figure}

The theoretical model presented in Sec.~\ref{sec:model} assumes that the laser is launched on the optical axis. This configuration is optimal for ELTs because it minimizes the spot elongation and the LGS aberrations. However, on a 10--meter class telescope, the laser can preferably be launched from the side of the telescope pupil, like for the Keck telescope~\citep{keck}. The LGS spots of a SH--WFS are not longer radially elongated from the pupil centre, but from an edge of the pupil. 

The model is still valid for this case, providing that the origin of the coordinate system ($\rho$,$\theta$) or ($x$,$y$) coincides with the position of laser launch telescope. Equation~\ref{equ:Z}, expressing the wavefront $Z(\rho,\theta)$, remains unchanged and the aberrations generated by a uniform thresholding are still centro--symmetric for a virtual giant pupil centred on the laser axis, with a diameter twice the distance between the laser axis and the optical axis. The only difference is the domain of definition of the function $Z(\rho,\theta)$, which must be shifted to match the telescope pupil. 

Figure~\ref{fig:offaxis} displays the portion of the wavefront $Z(\rho,\theta)$ seen by a circular pupil located at the right of the laser axis, and plots the peak--to--valley amplitudes of this wavefront projected on Zernike modes. Assuming a linear fit of the function $g(t)$, we can consider $Z(\rho)\propto\rho^2+8/3\,\rho^3$ for reproducing the LGS aberrations induced by an uniform thresholding (Eq.~\ref{equ:Ztri} with $E/\sigma=4$), and $Z(\rho)\propto\rho^2$ for reproducing the LGS aberrations induced by a radial thresholding (Eq.~\ref{equ:Zrho2}).    

As the aberrations are not centro--symmetric, the Zernike decomposition of $\rho^3$ features, in addition to the focus and spherical aberrations, many other modes, such as tip and tilt ($Z_2$, $Z_3$), astigmatism ($Z_5$, $Z_6$), coma ($Z_7$, $Z_8$) and some trefoil ($Z_9$, $Z_{10}$). Moreover, compared to the centrally projected laser configuration, the peak--to--valley amplitude of these aberrations are about 7 times greater.

On the other hand, the decomposition of $\rho^2$ features only tip and tilt modes in addition to the focus, meaning that the radial thresholding removes all the LGS aberrations beyond the focus. Tip--tilt and focus modes are not issues, since they will be corrected by the NGS tip--tilt and focus sensor already necessary to determine the atmospheric tip--tilt, unseen by a LGS WFS, and to offset the sodium layer distance variations, respectively.

Finally, the radial thresholding is even more advised if the laser is launched from the side of the pupil, since the LGS aberrations induced by the uniform thresholding are much higher and are spread in more modes. The side--projected laser configuration has no been tested on the UVic bench, but could be implemented if the need arises for any LGS AO project.

\section{Conclusion}
The fluctuations of the sodium layer vertical profile induce centro--symmetric and square symmetric aberrations on LGS SH--WFSs. Centro--symmetric aberrations, such as focus and spherical aberrations, are due to a spot truncation by either a circular field--stop or by a threshold applied on the image pixels. Square symmetric aberrations, such as $Z_{14}$, are due to a spot truncation by the pixel boundaries and a sampling effect. Some extra aberrations may arise due to uncalibrated non--common path errors of the LGS train varying with the sodium layer distance.

Both kind of aberrations have been reproduced in the laboratory with UVic LGS bench. Square symmetric aberrations should be negated with a polar coordinate CCD array. Considering such a detector and a field of view per sub--aperture wide enough to be able to image a 20--km thick sodium layer, the residual LGS aberrations are mainly caused by the threshold applied on the pixels of elongated spots prior to computing the centroids. 

This statement is confirmed by the theoretical model and experimental results presented in this paper. The aberrations induced by the thresholding contain focus and spherical aberration modes if the laser is launched from the pupil centre. Their amplitudes scale as the spot elongation (\ie the telescope diameter), the sodium profile asymmetry and the threshold value.  

These aberrations disappear if the threshold value is not static and uniform over the pupil, but defined dynamically and independently for each lenslet, proportionally to the maximum intensity of the local spot. The residual LGS aberrations are only a focus error if the laser is centrally projected, or a combination of tip, tilt and focus if the laser is launched beside the pupil. These residual modes are not issues since they will be corrected by the natural guide star tip--tilt and focus sensor already required in any LGS AO system. 

This new thresholding method, termed radial thresholding, is very simple and add no extra computational time compared to the uniform thresholding. It can be implemented on all current LGS SH--WFS in operation. Moreover, the radial thresholding makes the centre--of--gravity algorithm still well--suited for Extremely Large Telescopes, where the LGS spot elongation is greater. Unlike the matched filter or the correlation centroiding algorithms, the centre--of--gravity is a simple, well--proven and fast algorithm which requires no special calibration processes or reference images, and has no repercussions on the design of the LGS AO system.

\section*{Acknowledgments}
The authors are grateful to TMT consortium. The TMT Project gratefully acknowledges the support of the TMT partner institutions. They are the Association of Canadian Universities for Research in Astronomy (ACURA), the California Institute of Technology and the University of California. This work was supported as well by the Gordon and Betty Moore Foundation, the Canada Foundation for Innovation, the Ontario Ministry of Research and Innovation, the National Research Council of Canada, the Natural Sciences and Engineering Research Council of Canada, the British Columbia Knowledge Development Fund, the Association of Universities for Research in Astronomy (AURA) and the U.S. National Science Foundation. We would also like to thank Laurent Jolissaint, the reviewer of this paper, for his useful comments.



\begin{thebibliography}{99}

\bibitem[\protect\citeauthoryear{Davis et al.}{2006}]{davis}
Davis D. S., Hickson P., Herriot G., and She C-Y, 2006, Opt. Lett., 31, 3369

\bibitem[\protect\citeauthoryear{Clare et al.}{2007}]{clare}
Clare R. M.,van Dam M. A. and Bouchez A. H. , 2007, Optics Express, 15, 4711

\bibitem[\protect\citeauthoryear{Lardi\`ere et al.}{2008}]{lardiere}
Lardi\`ere O., Conan R., Bradley C., Jackson K. and Herriot G., 2008, Optics Express, 16, 5527

\bibitem[\protect\citeauthoryear{Poyneer}{2003}]{pyoneer}
Poyneer L. A., 2003, Applied Optics, 42, 5807

\bibitem[\protect\citeauthoryear{Thomas et al.}{2006}]{thomas}
Thomas S., Fusco T., Tokovinin A., Nicolle M., Michau V. and Rousset G., 2006,
MNRAS, 371, 323

\bibitem[\protect\citeauthoryear{Thomas et al.}{2008}]{thomas08}
Thomas S., Adkins S., Gavel D., Fusco T. and Michau V., 2008,
MNRAS, 387, 173

\bibitem[\protect\citeauthoryear{Gilles and Ellerbroek}{2008}]{gilles}
Gilles L. and Ellerbroek B., 2008, Opt. Lett. 33, 1159

\bibitem[\protect\citeauthoryear{Herriot et al.}{2006}]{herriot}
Herriot G., Hickson P., Ellerbroek B., V\'eran J.-P., Ye C. S., Clare R., and Looze D., 2006, Proc. SPIE, 6272, 62721I

\bibitem[\protect\citeauthoryear{Beletic et al.}{2005}]{beletic}
Beletic J. W., Adkins S., Burke B., Reich R., Kosicki B., Suntharalingham V., Bleau Ch., Duvarney R., Stover R., Nelson J. and Rigaut F., 2005, Experimental Astronomy, 19, 103

\bibitem[\protect\citeauthoryear{Ellerbroek et al.}{2008}]{tmt}
Ellerbroek B., Adkins S., Andersen D., et al., 2008, Proc. SPIE 7015, 70150R

\bibitem[\protect\citeauthoryear{Conan et al.}{2009}]{conan}
Conan R., Lardi\`ere O., Herriot G., Bradley C. and Jackson K., 2009, Applied Optics, 48, 1198





\bibitem[\protect\citeauthoryear{Noll}{1976}]{noll}
R. J. Noll, 1976, JOSA, 66, 207


\bibitem[\protect\citeauthoryear{Nicolle et al.}{2004}]{nicolle}
Nicolle M., Fusco T., Rousset G., and Michau V., 2004, Opt. Lett., 29, 2743 

\bibitem[\protect\citeauthoryear{Wizinowich et al.}{2006}]{keck}
Wizinowich P., Le Mignant D., Bouchez A., et al., 2006, PASP, 118, 297

\end{thebibliography}
\end{document}